\documentclass[floats,floatfix,showpacs,preprintnumbers,amssymb,prd,superscriptaddress,nofootinbib,nolongbibliography,reprint]{revtex4-1}
\usepackage[dvipdfmx]{graphicx}

\usepackage{subfigure}
\usepackage{bm}
\usepackage{mathrsfs}
\usepackage{amsmath}
\usepackage{cases}

\def\be{\begin{equation}}
\def\ee{\end{equation}}
\newcommand{\beq}{\begin{eqnarray}}
\newcommand{\eeq}{\end{eqnarray}} 
\usepackage{color}
\usepackage[normalem]{ulem}
\usepackage[linktocpage=true]{hyperref}
\usepackage{cancel}

\definecolor{jxgreen}{RGB}{0,150,0}

\definecolor{PPpurple}{RGB}{140,0,180}

\labelformat{section}{Section #1} 
\labelformat{subsection}{Section #1} 
\labelformat{subsubsection}{Section #1}
\labelformat{subsubsubsection}{Section #1}
\labelformat{equation}{Eq.~(#1)} 
\labelformat{figure}{Fig.~#1} 
\labelformat{subfigure}{Fig.~\thefigure#1} 
\labelformat{table}{Tab.~#1} 
\labelformat{appendix}{Appendix #1}

\newcommand{\sapienza}{Dipartimento di Fisica, Sapienza Università 
	di Roma, Piazzale Aldo Moro 5, 00185, Roma, Italy}
\newcommand{\infn}{INFN, Sezione di Roma, Piazzale Aldo Moro 2, 00185, Roma, Italy}

\begin{document}
\title{Vanishing Love Numbers of Kerr-Newman Black Holes}

\author{Sumanta Chakraborty}
\email{tpsc@iacs.res.in}
\affiliation{School of Physical Sciences, Indian Association for the Cultivation of Science, 2A Raja S.C. Mullick Road, Kolkata, 700032, India}

\author{Takuya Katagiri}
\email{takuya.katagiri@uniroma1.it}
\affiliation{\sapienza}\affiliation{\infn}

\author{Paolo~Pani}
\email{paolo.pani@uniroma1.it}
\affiliation{\sapienza}\affiliation{\infn}

\author{Jun-Xi Shi}
\email{shijunxi23@mails.ucas.ac.cn}
\affiliation{International Centre for Theoretical Physics Asia-Pacific,
University of Chinese Academy of Sciences, 100190 Beijing, China}
\affiliation{\sapienza}\affiliation{\infn}


\begin{abstract}
Tidal Love numbers characterize the conservative response of compact
objects to external perturbations and provide a powerful probe of
their nature. While the static Love numbers of Schwarzschild, Kerr,
and Reissner--Nordstr\"om black holes remarkably vanish, the tidal
response of a generic Kerr--Newman black hole has remained unexplored
due to the intrinsic coupling between gravitational and electromagnetic perturbations. Exploiting a recent breakthrough in the separability of these perturbations, we analytically determine the static tidal response of the most general stationary black hole in four-dimensional Einstein--Maxwell theory. For arbitrary multipole order, charge, and spin, we show that the entire matrix of conservative Love numbers vanishes, including the mixed gravito-electromagnetic couplings. We also compute the nonvanishing static dissipative response induced by frame dragging.
\end{abstract}

\maketitle

{\bf \em Introduction---}
The tidal response of a compact object provides a remarkably clean probe of its
internal structure and dynamics~\cite{Love1909,PoissonWill}. At linear order, its conservative response to a
slowly varying external field is encoded in tidal Love numbers~(LNs), which
relate the induced multipole moments to the applied tidal field (see~\cite{Chakraborty:2026qru,Rodriguez:2026iot} for recent reviews). Besides playing
an important role in the dynamics and gravitational-wave emission of compact
binaries, tidal LNs have emerged as powerful diagnostics of the nature of compact
objects and of the underlying theory of gravity
\cite{Cardoso:2017cfl,Chakraborty:2026qru,Rodriguez:2026iot}.

Black holes~(BHs) in four-dimensional vacuum General Relativity display a remarkable behavior in this respect. Their static LNs vanish for all bosonic perturbations, both for Schwarzschild and Kerr BHs~\cite{Damour_tidal, Binnington:2009bb, Damour:2009vw, Gurlebeck:2015xpa, Poisson:2014gka, Landry:2015zfa, Pani:2015hfa, LeTiec:2020bos, Chia:2020yla, LeTiec:2020spy, Bhatt:2023zsy}.
This property is highly nontrivial: generic material bodies, exotic compact objects, BHs in modified theories of gravity, in higher dimensions or surrounded by matter, and even BHs in four-dimensional General Relativity under fermionic perturbations generically exhibit a nonzero static tidal response~\cite{Cardoso:2017cfl, Chakraborty:2026qru, Chakraborty:2025zyb}.
The vanishing of static LNs for BHs in four-dimensional vacuum General Relativity has been linked to hidden symmetries of the perturbation equations~\cite{Hui:2021vcv,Charalambous:2022rre,Parra-Martinez:2025bcu} and their underlying hypergeometric structure~\cite{Ghosh:2026vig}. These connections suggest that the absence of a conservative static bosonic tidal response is not an accidental property of specific BH solutions, but rather reflects a deeper structure of linear perturbations in four-dimensional General Relativity.

Electric charge provides a particularly interesting test of this picture.
Already for the static Reissner--Nordstr\"om solution, gravitational and electromagnetic perturbations are coupled, so that the response is naturally described by a matrix of gravitational, electromagnetic, and mixed gravito-electromagnetic susceptibilities. Nevertheless, the corresponding static Love couplings vanish in four dimensions~\cite{Pereniguez:2021xcj, Rai:2024lho}. 
Remarkably, also in this case the vanishing can be traced to special symmetries of the static perturbation equations~\cite{Rai:2024lho}. 
This raises a natural question: does the same property of linear perturbations survive when \emph{both} charge and rotation are present in a BH spacetime?

The Kerr-Newman (KN) BH provides the decisive arena in which one can answer this question. 
It is the most general stationary, axi-symmetric and asymptotically flat BH solution of the four-dimensional Einstein-Maxwell theory, characterized by its mass, angular momentum, and electric charge~\cite{Newman:1965my}. Yet, despite its fundamental role in the space of BH solution in General relativity, the linear response of a generic KN BH has remained largely unexplored beyond its nonrotating (i.e., Reissner--Nordstr\"om) and neutral (i.e., Kerr) limits.
The main obstruction was a long-standing difficulty in BH perturbation theory: unlike for Kerr, the gravitational and electromagnetic perturbations of KN geometry are intrinsically coupled and, for generic rotation and charge, their exact separation into radial and angular equations had remained elusive. 
As a consequence, even the conceptually simple problem of determining the static tidal response of a generic KN BH has so far been inaccessible.

Very recently, this obstacle was overcome.
By reformulating the coupled Newman-Penrose equations and introducing a suitable matrix integrating factor, the full gravito-electromagnetic perturbation problem was shown to admit separation into radial and angular systems for generic sub-extremal KN BHs~\cite{Hintz:2026jiw}. 
This breakthrough enables the study of generic KN perturbations without relying on expansions in spin~\cite{Pani:2013ija, Pani:2013wsa} or charge~\cite{Mark:2014aja}, or computationally intensive solutions of coupled partial differential
equations~\cite{Dias:2015wqa, Dias:2022oqm}.
In particular, it allows the quasinormal-mode spectrum to be computed by solving ordinary differential equations~\cite{Destounis:2026ljd}.

In this Letter we exploit this separability to determine the static tidal response of KN BHs for the first time. 
We solve \emph{analytically} the separated equations in the zero-frequency limit, impose suitable boundary conditions at the future event horizon, and reconstruct the physical gravito-electromagnetic perturbations in order to identify their source and response pieces at spatial infinity. 
Due to the gravito-electromagnetic coupling, a gravitational tidal field could in principle induce an electromagnetic response and vice versa. 
However, we analytically show that both gravitationally and electromagnetically driven static LNs of the KN BH vanish identically, and determine the static dissipative response induced by frame dragging.

We have adopted geometrized Gaussian units, setting $G=c=4\pi\epsilon_0=1$.

{\bf \em Decoupled perturbation equations---}
The KN metric in Boyer--Lindquist coordinates reads~\cite{Newman:1965my}
\begin{align}
ds^2 ={}&
-\frac{\Delta}{\Sigma}
\left(dt-a\sin^2\theta\,d\phi\right)^2
+\frac{\Sigma}{\Delta}\,dr^2
+\Sigma\,d\theta^2
\nonumber\\
&+\frac{\sin^2\theta}{\Sigma}
\left[(r^2+a^2)d\phi-a\,dt\right]^2,
\end{align}
where $\Sigma=r^2+a^2\cos^2\theta$ and $\Delta=r^2-2Mr+a^2+Q^2$. Here $M$, $J=aM$, and $Q$ denote the BH mass, angular momentum, and electric charge, respectively, while $r_\pm=M\pm\sqrt{M^2-a^2-Q^2}$ are the two horizons. The electromagnetic potential reads $A_\mu dx^\mu = -\frac{Qr}{\Sigma}
\left(dt-a\sin^2\theta\,d\phi\right)$.

Following Hintz~\cite{Hintz:2026jiw}, we formulate linear gravito-electromagnetic perturbations in terms of suitable combinations of Newman–Penrose (NP) scalars~\cite{Newman:1961qr}, using the null Kinnersley tetrad $({\bm e_1},{\bm e_2},{\bm e_3},{\bm e_4})=({\bm \ell},{\bm n},{\bm m},\bar{\bm m})$~\cite{Kinnersley:1969zza}. The NP scalars are defined by
$\Psi_0=C_{1313}$, $\Psi_1=C_{1213}$, $\Psi_2=C_{1342}$,
$\Psi_3=C_{1242}$, $\Psi_4=C_{2424}$ and
$\Phi_0=F_{13}$, $\Phi_1=(F_{12}+F_{43})/2$, $\Phi_2=F_{42}$,
where $C_{abcd}$ and $F_{ab}$ denote tetrad components of the Weyl and electromagnetic field-strength tensors, respectively.
On the KN background, the only nonvanishing Weyl and Maxwell scalars are
$\Psi_2^{(0)}=(Q^2-M\bar\rho)/(\bar\rho\bar\rho^{*3})$ and
$\Phi_1^{(0)}=Q/(2\bar\rho^{*2})$, where
$\bar\rho=r+ia\cos\theta$ and $\bar{\rho}^*=r-ia \cos \theta$.

We introduce the combinations
\begin{equation}
\begin{aligned}
\label{eq:rescaledpsi}
\psi_2 &= \Psi_0^{(1)}, \quad
\psi_1=\sqrt{2}\bar{\rho}^*
\left(\Psi_1^{(1)}
-\frac{3\Psi_2^{(0)}}{2\Phi_1^{(0)}}\Phi_0^{(1)}\right),\\
 \psi_{-2}&= \left(\bar{\rho}^*\right)^4 \Psi_4^{(1)}, 
 \quad
\psi_{-1}=\frac{\left(\bar{\rho}^*\right)^3}{\sqrt{2}} \left(\Psi_3^{(1)} -\frac{3 \Psi_2^{(0)}}{2\Phi_1^{(0)}}\Phi_2^{(1)}\right),
\end{aligned}
\end{equation}
where the superscripts $(0)$ and $(1)$ denote background and perturbed quantities, respectively. These combinations are invariant under infinitesimal null rotations of the tetrad~\cite{Dias:2022oqm}.

Ref.~\cite{Hintz:2026jiw} casts the coupled gravito-electromagnetic perturbation equations into two first-order systems, labeled by $\varsigma=+1$ and $\varsigma=-1$ for the positive- and negative-spin sectors, respectively (see also Ref.~\cite{Motohashi:2026wac}). A matrix integrating factor then allows each system to be separated into radial and angular equations. Details of the separation and of the reconstruction of the physical fields are given in the Supplemental Material.

For extracting the tidal response, it is convenient to work with the radiative scalars $\Psi_0^{(1)}$ and $\Phi_0^{(1)}$. While Ref.~\cite{Hintz:2026jiw} employs the phantom gauge, in which $\Phi_0^{(1)}=0$, we instead perform a null rotation such that $\Psi_1^{(1)}=0$. In this frame, $\Phi_0^{(1)}=-(\sqrt{2}\Phi_1^{(0)}/3\bar{\rho}^*\Psi_2^{(0)})\psi_1$ is directly related to $\psi_1$, while $\Psi_0^{(1)}=\psi_2$, so that the tidal response can be extracted from $\psi_1$ and $\psi_2$. Finally, the variables $\psi_{\pm1}$ and $\psi_{\pm2}$ can be reconstructed as linear combinations of a master radial function and its radial derivative, thereby reducing the perturbation problem to a single master function.

{\bf \em Static perturbations---}
In the static ($\omega\to0$) limit of interest here, the radial master function $R_{\ell m,b}^{(\varsigma)}$ satisfies the following differential equation,
\begin{align}\label{radial}
\dfrac{d^{2}R_{\ell m,b}^{(\varsigma)}}{dr^{2}}&+\left[\frac{(1-3\varsigma)\Delta'}{2\Delta}+\frac{2iam}{\Delta}\right]\dfrac{dR_{\ell m,b}^{(\varsigma)}}{dr}
\nonumber
\\
&+\left[\frac{9M^{2}}{4Q^{2}}-\lambda_{\ell,b}^{2}\right]\frac{R_{\ell m,b}^{(\varsigma)}}{\Delta}=0\,, 
\end{align}
Remarkably, the angular equations depend on $a$ and $\omega$ only through $a\omega$ (see Supplemental Material). Thus, in the static limit the angular eigenfunctions are spin-weighted spherical harmonics, despite the rotation of the background, and the eigenvalue problem can be solved analytically for arbitrary $(Q,a)$, yielding two branches:
\begin{equation}\label{ang_eigen}
\lambda\equiv\lambda_{\ell,b}=b\sqrt{\left(\frac{3M}{2Q}\right)^2+(\ell-1)(\ell+2)}\,,
\end{equation}
where $b=\pm1$ denotes these branches, while $(\ell,m)$ are integers satisfying $|m|\leq\ell$ and $\ell\geq2$, for the paired gravito-electromagnetic sector considered here. Upon substituting the angular eigenvalue of~\ref{ang_eigen} into~\ref{radial}, the apparent singularities of the radial equation and of $\lambda$ in the $Q\to0$ limit cancel, consistently with the numerical results in the dynamical case~\cite{Destounis:2026ljd}. Interestingly, in the static limit the radial equation depends on the angular eigenvalue $\lambda_{\ell,b}$ only through $\lambda_{\ell,b}^2$ and is therefore insensitive to its sign (or, equivalently, the radial equation does not depend on the choice of $b$). Consequently, the gravitational-led ($b=+1$) and electromagnetic-led ($b=-1$) perturbations~\cite{Destounis:2026ljd} obey the same radial equation, despite belonging to distinct perturbation sectors. 

The general solution to~\ref{radial} can be expressed in closed form in terms of hypergeometric functions as
\begin{align}
&R_{\ell m,b}^{(\varsigma)}=c_{1,b}\,_{2}F_{1}(\alpha_+,\alpha_-;\gamma;-z)
\nonumber
\\
&+c_{2,b}(-z)^{1-\gamma}\,_{2}F_{1}(\alpha_+-\gamma+1,\alpha_--\gamma+1;2-\gamma;-z)\,,
\label{eq:general-solution}
\end{align}
where $z\equiv (r-r_{+})/(r_{+}-r_{-})$, $2\alpha_\pm=\pm{(1+2\ell)}-{3\varsigma}$, $\gamma={(1-3\varsigma)}/{2}+p$, and $p={2iam}/(r_{+}-r_{-})$.
The above solution is valid for $am\neq0$; the case $am=0$ is separately discussed in the Supplemental Material.

{\bf \em Love matrix of KN BHs---}
Without loss of generality, we focus on the positive-spin sector, $\varsigma=1$, and on the gauge-invariant radiative degrees of freedom $\psi_{1,2}$.

Due to the gravito-electromagnetic coupling, $\psi_{1,2}$, depend on both gravitational-led and electromagnetic-led perturbations, through a linear combination of $R_{\ell m,\pm}^{(+1)}$ and their radial derivatives. 
As detailed in the Supplemental Material, the asymptotic behavior of a $(\ell,m)$ mode schematically reads
\begin{align}
\psi_{2,\ell m}
\simeq r^{\ell-2}& \Bigl[
A^{\rm g}_{\ell m}(1+\cdots)
\notag\\
&\qquad
+\dfrac{B^{\rm g}_{\ell m}}{r^{2\ell+1}}(1+\cdots)
\Bigr]{}_2S_{\ell m}(\theta)e^{im\phi},
\label{eq:reduced-radial-equation}
\\
\psi_{1,\ell m}
\simeq r^{\ell-1}& \Bigl[
A^{\rm e}_{\ell m}(1+\cdots)
\notag\\
&\qquad
+\dfrac{B^{\rm e}_{\ell m}}{r^{2\ell+1}}(1+\cdots)
\Bigr]{}_1S_{\ell m}(\theta)e^{im\phi}\,,
\end{align}
where the dots denote sub-leading terms,  ${}_sS_{\ell m}(\theta)$ are the spin-weight-$s$ spherical harmonics, whereas the four coefficients $A_{\ell m}^{g,e}$ and $B_{\ell m}^{g,e}$ are linear combinations of $c_{1,\pm}$ and $c_{2,\pm}$.

To extract the behavior near the horizon, $r\sim r_+$, it is convenient to study the dynamical system in ingoing null coordinates, such that the radial function in the static limit behaves as~\cite{Teukolsky:1974yv},
\begin{align}\label{psi0boundary}
    \psi_s &\sim \frac{{\cal A}_{\rm ingoing}^{(s)}}{\Delta^s}+{\cal A}_{\rm outgoing}^{(s)}e^{2i\bar\omega r_*} 
    \nonumber
    \\
    &\to \frac{{\cal A}_{\rm ingoing}^{(s)}}{\Delta^s}+{\cal A}_{\rm outgoing}^{(s)}z^{-p}\,,
\end{align}
for $s=1,2$, where $dr/dr_*=\Delta/(r^2+a^2)$ defines the tortoise coordinate, $\bar \omega \equiv \omega -m\Omega_H$, and $\Omega_H=a/(r_+^2+a^2)$ is the horizon angular velocity.
The amplitudes of the ingoing and outgoing modes are also linear combinations of $c_{1,\pm}$ and $c_{2,\pm}$.
Imposing purely ingoing waves at the future horizon, we set ${\cal A}_{\rm outgoing}^{(s)}=0$ (for $s=1,2$), corresponding to $c_{2,\pm}=0$ in~\ref{eq:general-solution}.
Interestingly, this condition also ensures that the radial master functions $R^{(+1)}_{\ell m,b}$ are smooth at the future event horizon. At large radius, they behave as
\begin{equation}
\label{eq:Ratlarger}
R^{(+1)}_{\ell m,b}
=\mathcal N_{\ell m,b} r^{\ell+2}
\Bigg[(1+\cdots)+\mathcal F_{\ell m}(1+\cdots)
\left(\frac{r_+}{r}\right)^{2\ell+1}
\Bigg]\,,
\end{equation}
where $\mathcal N_{\ell m,b}$ are generic amplitudes, and
\begin{equation}
\begin{aligned}
\mathcal F_{\ell m}
=
&-\frac{iam}{r_+}
\frac{(\ell+2)!(\ell-2)!}
{(2\ell+1)!(2\ell)!}\times\\
&\prod_{j=1}^{\ell}
\left[
j^2\left(1-\frac{r_-}{r_+}\right)^2
+\left(\frac{2am}{r_+}\right)^2
\right]\,,
\end{aligned}
\end{equation}
is independent of $b=\pm1$. As noted earlier, this expression holds for $\ell\ge 2$. The $\ell=1$ case will be discussed separately.

In terms of the asymptotic amplitudes, the purely ingoing boundary condition at the horizon for $\psi_{1,2}$ (or, regularity condition at the horizon for $R^{(+1)}_{\ell m,b}$) can be written in matrix form as a relation between $B_{\ell m}^{g,e}$ and $A_{\ell m}^{g,e}$
\begin{equation}
    \binom{B_{\ell m}^{\rm  g}}{B_{\ell m}^{\rm  e}}
=
\mathsf F_{\ell m}r_+^{2\ell+1}
\binom{A_{\ell m}^{\rm  g}}{A_{\ell m}^{\rm  e}}.
\end{equation}
where we define the \emph{Love matrix},
\begin{widetext}
\begin{equation}
    \mathsf F_{\ell m}
=
\begin{pmatrix}
\mathcal F_{\ell m}^{\rm gg}&\mathcal F_{\ell m}^{\rm ge}\\
\mathcal F_{\ell m}^{\rm eg}&\mathcal F_{\ell m}^{\rm ee} \label{LoveMatrix}
\end{pmatrix}
=
\mathcal F_{\ell m}\frac{(2\ell+1)(\ell-1)}{3(M^2-Q^2)}
\begin{pmatrix}
\dfrac{3M^2+2(\ell-1)Q^2}{(2\ell+1)(\ell-1)}
&
\dfrac{-Q^2}{\sqrt{(\ell-1)(\ell+2)}}
\\
\dfrac{-Q^2}{\sqrt{(\ell-1)(\ell+2)}}
&
\;\;
\dfrac{2(\ell+2)Q^2-3M^2}{(\ell+2)(2\ell+1)}
\end{pmatrix}.
\end{equation}
\end{widetext}
In the $Q\to 0$ limit, we have 
$\mathsf F_{\ell m}=\mathcal{F}_{\ell m}\,{\rm diag}\left\{1,-(\ell-1)/(\ell+2)\right\}$, whose components respectively reduce to the purely gravitational ($s=\pm2$) and electromagnetic ($s=\pm1$) tidal response of a Kerr BH~\cite{Chia:2020yla,Chakraborty:2026qru}.

This analytical expression, valid for a generic KN BH, is the main result of this work and warrants a few comments.
First, the Love matrix is obtained by first analytically continuing for arbitrary $(\ell,m)$ and only at the end taking the $(\ell,m)\in \mathbb{Z}$ limit. This allows us to unambiguously identify the response coefficients through analytic continuation of the angular momentum index $\ell$~\cite{Kol:2011vg,LeTiec:2020bos}.
Furthermore, as expected, the mixed coefficients are symmetric, ${\cal F}_{\ell m}^{\rm ge}={\cal F}_{\ell m}^{\rm eg}$, and vanish in the $Q\to0$ limit, in which the gravitational and electromagnetic sectors decouple.
Most importantly, all entries of the response matrix are purely \emph{imaginary} and proportional to $am$, as in the Kerr case~\cite{Chia:2020yla,LeTiec:2020bos,LeTiec:2020spy}. Therefore, the conservative static tidal response vanishes identically, while the nonvanishing imaginary part describes the dissipative response associated with frame dragging~\cite{Chakraborty:2025zyb}.

This result is remarkable in the generic KN case. In principle, the conservative response could have contained mixed contributions proportional, for example, to $aQ^2$, as those appearing in the dissipative part of the mixed couplings. Those terms vanish both in the Kerr ($Q=0$) and Reissner–Nordstr\"om ($a=0$) limits while remaining finite when both charge and rotation are present. Our result demonstrates that no such terms arise: the conservative tidal response continues to vanish even when both spin and charge are simultaneously nonzero.

Strictly speaking, \ref{LoveMatrix} was derived for $am\neq0$. However, in the Supplemental Material we show that, when $am=0$, $\mathsf F_{\ell m}$ identically vanishes, consistently with the above result.

Finally, in the Supplemental Material we also analyze the electric dipole mode with $\ell=1$. In this case, the gravitational sector is nondynamical and the response is purely electromagnetic,
\begin{align}
    \mathcal{F}^{\rm ee}_{1m} &\equiv \frac{B^{\rm e}_{1m}}{r_+^3 A^{\rm e}_{1m}} \nonumber\\
    &=i \frac{2am}{3r_{+}}\left(\frac{M^{2}-2Q^{2}}{M^{2}-Q^{2}}\right)
\frac{
M^2-Q^2+\left(m^2-1\right)a^2
}{r_+^2}\,,
\end{align}
sharing the same qualitative features as the general result. The neutral limit, $Q\to0$, reproduces the Kerr result~\cite{Chakraborty:2026qru}.

{\bf \em Discussion---}
We have determined analytically the static tidal response of generic KN BHs, which constitute the most general four-dimensional, asymptotically flat electrovacuum BH family. Our analysis exploits the recently established separability of the coupled gravito-electromagnetic perturbation equations of a KN BH~\cite{Hintz:2026jiw}. Remarkably, in the static limit both the angular and radial equations admit analytic solutions for arbitrary charge, spin, and multipole indices $(\ell,m)$, allowing us to impose suitable boundary conditions at the future event horizon and extract the tidal response at spatial infinity.

Our main result is that the entire conservative Love matrix vanishes identically, extending the remarkable rigidity of Schwarzschild, Kerr, and Reissner--Nordstr\"om BHs to the full KN family, under static tidal perturbations. This result is particularly nontrivial because gravitational and electromagnetic perturbations are intrinsically coupled, so that the tidal response contains purely gravitational, purely electromagnetic, and mixed gravito-electromagnetic susceptibilities.

The absence of conservative tidal deformability does not, however, imply the absence of a tidal response. For nonaxisymmetric static perturbations, the response has a nonvanishing imaginary part, associated with a dissipative tidal torque. Although the external field is stationary at infinity, it varies along the horizon generators with frequency $\bar\omega=\omega-m\Omega_H$, which remains nonzero as $\omega\to0$ for $m\Omega_H\neq0$. A strictly stationary tide can therefore transfer angular momentum and increase the horizon area while the energy flux associated with the asymptotic time translation vanishes. This response is the zero-frequency limit of the dissipative response responsible for absorption and superradiance at finite frequency~\cite{Brito:2015oca}.

The persistence of this property in the most general asymptotically flat electrovacuum BH solution points to a deeper structure of four-dimensional Einstein--Maxwell theory, possibly related to hidden symmetries of the perturbation equations~\cite{Hui:2021vcv,Charalambous:2022rre,Parra-Martinez:2025bcu}.

Indeed, our result can also be traced to a simple ladder structure of the static radial equation. For $\varsigma=1$, introducing $x=2(r-M)/(r_+-r_-)$,~\ref{radial} reduces to the Jacobi equation, with horizon-regular solution
\begin{equation}
R_{\ell m,b}\propto P_{\ell+2}^{(-2+p,-2-p)}(x)\,.
\end{equation}
Since the Jacobi indices are independent of $\ell$, the corresponding recurrence relations naturally define a ladder structure connecting neighboring
multipoles $\ell\to\ell\pm1$. 
For integer $\ell$, the horizon-regular solution for the radial master function truncates to a Jacobi polynomial, providing a structural explanation for the absence of a conservative tidal response, namely enforcing ${\rm Re}\left[{\cal F}_{\ell m}\right]=0$. Since regularity of the master functions is equivalent to imposing ingoing-wave boundary conditions on the reconstructed physical fields $\psi_{1,2}$, this ladder structure directly accounts for the vanishing of the entire conservative Love matrix.
Moreover, the hypergeometric form of the static radial solution, \ref{eq:general-solution}, directly exhibits a ladder symmetry~\cite{Ghosh:2026vig}.
Although a more detailed analysis of the underlying symmetry structure is warranted, this observation suggests that the ladder structure previously identified for simpler BH backgrounds~\cite{Hui:2021vcv,Rai:2024lho,Katagiri:2022vyz} extends to generic KN BHs as well. 

Natural extensions of our work include computing the dynamical LNs~\cite{Saketh:2023bul,Katagiri:2024wbg, Chakraborty:2025wvs,Combaluzier--Szteinsznaider:2025eoc,Chakraborty:2026qru,Kosmopoulos:2025rfj} of KN BHs, as well as matching the tidal response to an effective field theory description, as previously done for Kerr and Reissner–Nordstr\"om BHs. It would be interesting to investigate whether the vanishing of all conservative gravito-electromagnetic Love couplings can be understood in terms of hidden symmetries of the KN perturbation equations. Further directions include extending the separability framework to fermionic perturbations, whose static tidal response is expected to be nonvanishing, and exploring the tidal response beyond linear perturbation theory.

\noindent
{\bf \em  Acknowledgments---}
We acknowledge financial support by the MUR FIS2 Advanced Grant ET-NOW (CUP:~B53C25001080001) and by the INFN TEONGRAV initiative. 
The research of SC is supported by MATRICS (MTR/2023/000049) and Core Research Grants (CRG/2023/000934) from SERB, ANRF, Government of India. 
J.~S. acknowledges financial support from the China Scholarship Council (CSC) for joint PhD training in Rome.

\bibliography{Ref}

\newpage

\appendix

%
\section*{Supplemental Material}
Here we provide technical details of the analysis presented in the main text, focusing in particular on the extraction of the tidal response from NP scalars reconstructed from the separated variables.

\subsection{Tetrad frame choice}
\label{sec:nulltetradrotations}
We first discuss the freedom in the choice of tetrad under null rotations. Let $\beta_+$ and $\beta_-$ be infinitesimal complex-valued functions parametrizing null rotations that leave ${\bm n}$ and ${\bm \ell}$ fixed, respectively. To first order, the corresponding transformations are
$
{\bm n}'={\bm n},~{\bm m}'={\bm m}+\beta_+{\bm n},~{\bm \ell}'={\bm \ell}+\bar{\beta}_+{\bm m}+\beta_+\bar{\bm m}
$
and
$
{\bm \ell}'={\bm \ell},~{\bm m}'={\bm m}+\bar{\beta}_-{\bm \ell},~{\bm n}'={\bm n}+\beta_-{\bm m}+\bar{\beta}_-\bar{\bm m}.
$
On the KN background, the induced first-order transformations of the NP scalars are~\cite{Hintz:2026jiw,Motohashi:2026wac}
\begin{equation}
\begin{aligned}
\label{eq:transformationPhi0Psi1}
 \Phi_0^{(1)'} =\Phi_0^{(1)}+ 2\beta_+ \Phi_1^{(0)},&\quad    \Psi_1^{(1)'}= \Psi_1^{(1)}+3 \beta_+ \Psi_2^{(0)},\\
 \Phi_2^{(1)'}= \Phi_2^{(1)}+2\beta_-\Phi_1^{(0)},& \quad \Psi_3^{(1)'} = \Psi_3^{(1)} + 3\beta_- \Psi_2^{(0)}.
\end{aligned}
\end{equation}

This equation shows that $\beta_+$ can be chosen to set either $\Phi_0^{(1)'}$ or $\Psi_1^{(1)'}$ to zero, while $\beta_-$ can similarly be chosen to set either $\Phi_2^{(1)'}$ or $\Psi_3^{(1)'}$ to zero. Specifically, for $\Phi_1^{(0)}\neq0$, choosing 
\begin{align}
\label{eq:phantomgauge}
    \beta_+= -\frac{\Phi_0^{(1)}}{2\Phi_1^{(0)}},\quad \beta_-=-\frac{\Phi_2^{(1)}}{2\Phi_1^{(0)}},
\end{align}
sets $\Phi_0^{(1)'}=\Phi_2^{(1)'}=0$. This choice is called the phantom gauge~\cite{ChandraBook,Hintz:2026jiw}. The corresponding Weyl scalars are
\begin{equation}
\begin{aligned}
    \Psi_1^{(1)'} = & \Psi_1^{(1)}-  \frac{3 \Psi_2^{(0)} }{2\Phi_1^{(0)}} \Phi_0^{(1)},\quad
    \Psi_3^{(1)'} =  \Psi_3^{(1)}- \frac{3 \Psi_2^{(0)}}{2\Phi_1^{(0)}}  \Phi_2^{(1)}.
\end{aligned}
\end{equation}
Alternatively, for $\Psi_2^{(0)}\neq 0$, choosing
\begin{align}
\label{eq:transversegauge}
    \beta_+=- \frac{\Psi_1^{(1)}}{3\Psi_2^{(0)}},\quad \beta_-=- \frac{\Psi_3^{(1)}}{3\Psi_2^{(0)}},
\end{align}
sets $\Psi_1^{(1)'}=\Psi_3^{(1)'}= 0$, yielding
\begin{equation}
\begin{aligned}
\Phi_0^{(1)'}=& \Phi_0^{(1)}-\frac{2\Phi_1^{(0)}}{3\Psi_2^{(0)}}\Psi_1^{(1)},\quad
\Phi_2^{(1)'}= \Phi_2^{(1)}-\frac{2\Phi_1^{(0)}}{3\Psi_2^{(0)}}\Psi_3^{(1)}. 
\end{aligned} 
\end{equation}

The rescaled NP scalars defined in~\ref{eq:rescaledpsi} of the main text are invariant under null rotations, whereas $\Psi_1^{(1)}$, $\Psi_3^{(1)}$, $\Phi_0^{(1)}$, and $\Phi_2^{(1)}$ individually depend on the choice of tetrad frame. We adopt the frame that remains well defined in the uncharged limit $Q\to0$, obtained by imposing $\Psi_1^{(1)'}=\Psi_3^{(1)'}=0$ through the transformation in~\ref{eq:transversegauge}.
The resulting expressions are then 
\begin{align}
\label{eq:Phi01_in_psi1}
    \psi_1 = -\bar{\rho}^* \frac{3\Psi_2^{(0)}}{\sqrt{2}\Phi_1^{(0)}}  \Phi_0^{(1)'},\quad \psi_{-1}=-\left(\bar{\rho}^*\right)^3 \frac{3\Psi_2^{(0)}}{2\sqrt{2}\Phi_1^{(0)}}\Phi_2^{(1)'},
\end{align}
while $\psi_2=\Psi_0^{(1)}$ and $\psi_{-2}=(\bar\rho^*)^4\Psi_4^{(1)}$ are unaffected by the rotation.
Note that $\Phi_1^{(0)}\propto Q$, while $\Psi_2^{(0)}$ remains finite and nonzero as $Q\to 0$. The Maxwell scalars have finite uncharged limits when $Q\psi_1$ and $Q\psi_{-1}$ have finite limits.

\subsection{Decoupled perturbation equations}
\label{sec:DecoupledEquations}
Here we review the separation procedure of Ref.~\cite{Hintz:2026jiw}. The linearized NP equations yield two sets of four coupled first-order differential equations for the rescaled NP scalars and spin coefficients : $(\psi_{+2},\psi_{+1},\hat{k},\hat{s})$ for the positive-spin system $\varsigma=+1$, and $(\psi_{-2},\psi_{-1},\hat{n},\hat{\ell})$ for the negative-spin system $\varsigma=-1$.
Although Ref.~\cite{Hintz:2026jiw} adopts the phantom gauge, the first-order systems and the subsequent derivation can be formulated entirely in terms of quantities invariant under null rotations.
These systems can be cast into
\begin{align}
\label{eq:eqforzpm_decoupled}
    \left[\Delta^{1/2} \partial_r - \left(R_0^{(\varsigma)}(r) {\cal I}+R_1^{(\varsigma)}(r) {\cal J} \right)-{\cal A}^{(\varsigma)}(\theta)\right]
    \begin{pmatrix}
        z_+^{(\varsigma)}\\ z_-^{(\varsigma)}
    \end{pmatrix}
    =0,
\end{align}
where the radial functions~$R_0^{(\varsigma)}(r)$ and $R_1^{(\varsigma)}(r)$ are given in Ref.~\cite{Hintz:2026jiw}. Each $z_\pm^{(\varsigma)}$ is a two-component vector, defined through 
\begin{equation}
    \begin{aligned}
\label{eq:zpzmp}
    z_+^{(+1)} & := \Delta^{5/4} \sqrt{\sin \theta} \left({\rm det} X\right) X^{-1}  \begin{pmatrix}
        \psi_2 + i \Delta^{-1/2}\psi_1\\
        -2Q \left(\hat{k}+ i \Delta^{-1/2} \hat{s}\right)
    \end{pmatrix},\\
    z_-^{(+1)}&:= \Delta^{5/4} \sqrt{\sin \theta}X^* \begin{pmatrix}
        \psi_2 - i \Delta^{-1/2}\psi_1\\
        -2Q \left( \hat{k} - i \Delta^{-1/2} \hat{s}\right)
    \end{pmatrix}
    \end{aligned}
\end{equation}
and
\begin{equation}
    \begin{aligned}
    \label{eq:zpzmm}
    z_+^{(-1)} & := \Delta^{-3/4} \sqrt{\sin \theta} \left({\rm det} X\right) X^{-1}  \begin{pmatrix}
        \psi_{-2} - i \Delta^{1/2}\psi_{-1}\\
        2Q \left(\hat{n}- i \Delta^{1/2}\hat{\ell}\right)
    \end{pmatrix},\\
    z_-^{(-1)}&:= \Delta^{-3/4} \sqrt{\sin \theta}  X^* \begin{pmatrix}
        \psi_{-2} + i \Delta^{1/2}\psi_{-1}\\
        2Q \left(\hat{n} + i \Delta^{1/2} \hat{\ell}\right)
    \end{pmatrix},
\end{aligned}
\end{equation}
with
\begin{align}
    X=\begin{pmatrix}
        Q \dfrac{\Delta^{1/2}+a\sin\theta}{\bar{\rho}} & -i \left(M-\dfrac{Q^2}{\bar{\rho}}\right)\\
        i \left(M-\dfrac{Q^2}{\bar{\rho}^*}\right) & Q\dfrac{\Delta^{1/2}-a\sin \theta}{\bar{\rho}^*}
    \end{pmatrix}.
\end{align}
The constant matrices~${\cal I}$ and ${\cal J}$ are defined by
\begin{align}
    {\cal I}=
    \begin{pmatrix}
       {\bf 0} &I_2\\
       I_2  & {\bf 0}
    \end{pmatrix},
    \quad
    {\cal J} = \begin{pmatrix}
        {\bf 0} & -J\\
        J & {\bf 0}
    \end{pmatrix},
    \quad 
    J= \begin{pmatrix}
        0 &-1\\
        1 & 0
    \end{pmatrix},
\end{align}
where $I_N$ denotes the $N\times N$ identity matrix. The~$4\times 4$ matrix differential operator~${\cal A}^{(\varsigma)}$ is defined by
\begin{align}
    {\cal A}^{(\varsigma)}:= \begin{pmatrix}
        i I_{2}  \partial_\theta & i \Theta^{(\varsigma)}\\
        -i \Theta^{(\varsigma)} & -i I_{2} \partial_\theta
    \end{pmatrix},\quad 
        \Theta^{(\varsigma)} (\theta)= 
    \begin{pmatrix}
        \Theta_0^{(\varsigma)} & \Theta_1^{(\varsigma)}\\
        \Theta_1^{(\varsigma)} & -\Theta_0^{(\varsigma)}
    \end{pmatrix},
\end{align}
and the components of $\Theta^{(\varsigma)}(\theta)$ are explicitly given in Ref.~\cite{Hintz:2026jiw}.

We now introduce the separated ansatz for a single Fourier mode with frequency $\omega$ and azimuthal number $m$:
\begin{equation}
    \begin{aligned}
\label{eq:zpzmDecomposition}
    \begin{pmatrix}
        z_+^{(\varsigma)} \\ z_-^{(\varsigma)}
    \end{pmatrix}
    =& \left[ z_1^{(\varsigma)} (r) \left(u(\theta)+v(\theta) \right) - i z_2^{(\varsigma)}(r) \left(u(\theta)-v(\theta)\right)\right]\\
    &\times e^{-i \omega t+i m\phi},
\end{aligned}
\end{equation}
where the four-component vectors~$u(\theta)$ and $v(\theta)$ satisfy
\begin{align}
\label{eq:eigeneqs}
    {\cal A}^{(\varsigma)} u= \lambda u,\quad {\cal S} u = i u, \quad v:=  {\cal I} u.
\end{align}
Here, $\lambda$ is the separation eigenvalue and ${\cal S}:= {\cal IJ}$. We discuss the angular eigenvalue problem in the next section. The scalar radial functions~$z_1^{(\varsigma)}$ and $z_2^{(\varsigma)}$ are expressed in terms of a radial master function~$R^{(\varsigma)}(r)$ as
\begin{equation}
    \begin{aligned}
    \label{eq:eaforz12p}
  z_1^{(+1)}& = \left(R_1^{(+1)}+i \lambda \right)^{-1} \Delta^{-1/4} H e^{-2 i \omega Q^2r/(M^2-Q^2)}\partial_r R^{(+1)} ,\\
   z_2^{(+1)}&=\Delta^{-3/4}H  e^{- 2 i \omega Q^2r/(M^2-Q^2)} R^{(+1)},
\end{aligned}
\end{equation}
and 
\begin{equation}
    \begin{aligned}
     \label{eq:eaforz12m}
  z_1^{(-1)}& = \Delta^{3/4}H  e^{- 2 i \omega Q^2r/(M^2-Q^2)} R^{(-1)},\\
   z_2^{(-1)}&= \left(R_1^{(-1)}-i \lambda \right)^{-1} \Delta^{5/4} H e^{-2 i \omega Q^2r/(M^2-Q^2)}\partial_r R^{(-1)},
\end{aligned}
\end{equation}
with $H = \exp\left(i \int_{2r_+}^rdr {K}/{\Delta}\right)$ and $K=-(r^2+a^2)\omega+am$. Substituting this ansatz into~\ref{eq:eqforzpm_decoupled} and using~\ref{eq:eigeneqs}, one obtains the following second-order radial equation:~\cite{Hintz:2026jiw}
\begin{widetext}
    \begin{align}
\label{eq:eqforRvarsigma}
    \left[ \partial_r - \left(\left(1+3\varsigma\right) \frac{\partial_r \Delta}{2 \Delta}-\frac{2i \tilde{K}}{\Delta}+\frac{\partial_r R_1^{(\varsigma)}}{R_1^{(\varsigma)}+i \varsigma \lambda }\right) \right]\Delta \partial_r R^{(\varsigma)} -\left[\left(R_1^{(\varsigma)}\right)^2+\lambda^2 \right]R^{(\varsigma)}=0\,,
\end{align}
\end{widetext}
where $\tilde{K}:= K-2\omega Q^2\Delta/(M^2-Q^2)$. In the static limit~($\omega\to0$),~\ref{eq:eqforRvarsigma} reduces to~\ref{radial} in the main text. 

\subsection{Angular eigenvalue problem in the static limit}
\label{sec:AngularSolution}
We determine the regular angular eigenfunctions in the zero-frequency limit. Since the angular equation depends on $\omega$ and $a$ only through the combination $a\omega$~\cite{Hintz:2026jiw}, the static angular problem reduces to its nonrotating form.

The eigenfunctions satisfy~\ref{eq:eigeneqs}, following which, we parametrize the eigenvectors as
\begin{equation}
\label{eq:angulareigenfunctions}
    u=\frac12
\begin{pmatrix}
iF+G\\
F-iG\\
-iF+G\\
F+iG
\end{pmatrix},
\qquad
v=\frac12
\begin{pmatrix}
-iF+G\\
F+iG\\
iF+G\\
F-iG
\end{pmatrix}.
\end{equation}
In the static limit~$\omega \to 0$,~\ref{eq:eigeneqs} then reduces to
\begin{align}
\left(\partial_\theta+\Theta_0^{(\varsigma)}\right)G
&=\left(\lambda+\frac{3M}{2Q}\right)F,
\\
\left(-\partial_\theta+\Theta_0^{(\varsigma)}\right)F
&=\left(\lambda-\frac{3M}{2Q}\right)G,
\label{eq:reduced-eigenfunction}
\end{align}
where
\[
\Theta_1^{(\varsigma)}=-\frac{3M}{2Q},
\,\;
\Theta_0^{(\varsigma)}
=\frac32\cot\theta+\frac{\mathfrak{m}}{\sin\theta},
\,\;
\mathfrak{m}:=\varsigma m\,.
\]
Eliminating $G$ and defining
$F=C^{\rm e}  \sqrt{\sin\theta}\,S$, we obtain from \ref{eq:reduced-eigenfunction},
\begin{equation}
\frac{\partial_\theta
\left(\sin\theta\,\partial_\theta S\right)}{\sin\theta}
+
\left[
\lambda^2-\left(\frac{3M}{2Q}\right)^2+1
-\frac{(\mathfrak{m}+\cos\theta)^2}{\sin^2\theta}
\right]S=0.
\label{eq:polar-spherical}
\end{equation}
This is the angular equation for spin-weight-$1$ spherical harmonics.
Regularity therefore yields
\begin{equation}
\lambda_{\ell,b}
=
b\sqrt{
\left(\frac{3M}{2Q}\right)^2
+(\ell-1)(\ell+2)},
\qquad b=\pm1 \,,
\end{equation}
where $\ell$ is an integer.
The two signs define the
gravitational-led ($b=+1$) and electromagnetic-led ($b=-1$)
branches, as in the dynamical case~\cite{Destounis:2026ljd}. Indeed, as shown for static perturbations in~\ref{sec:NPreconstruction}, in the uncharged limit~$(Q\to 0)$, the $b=+1$ and $b=-1$ branches reduce to purely gravitational and purely electromagnetic perturbations, respectively. 

Using the standard spin-raising relation, the two angular functions
can be written as
\begin{equation}
\label{eq:solsFG}
F_{\ell\mathfrak m,b}
=C_{\ell ,b}^{\rm e}\sqrt{\sin\theta}\,
{}_1S_{\ell\mathfrak m},
\qquad
G_{\ell\mathfrak m,b}
=C_{\ell ,b}^{\rm g}\sqrt{\sin\theta}\,
{}_2S_{\ell\mathfrak m},
\end{equation}
where ${}_sS_{\ell\mathfrak{m}}(\theta)$ are spin-weight-$s$ spherical harmonics, and
\begin{equation}
\begin{aligned}
\label{eq:CfCg}
C_{\ell ,b}^{\rm e}
&=
\frac{\sqrt{(\ell-1)(\ell+2)}}
{\sqrt{(\ell-1)(\ell+2)+(\lambda_{\ell,b}+3M/2Q)^2}},
\\
C_{\ell ,b}^{\rm g}
&=
\frac{\lambda_{\ell,b}+3M/(2Q)}
{\sqrt{(\ell-1)(\ell+2)+(\lambda_{\ell,b}+3M/2Q)^2}}.
\end{aligned}
\end{equation}
The coefficients satisfy $|C_{\ell ,b}^{\rm e}|^2+|C_{\ell ,b}^{\rm g}|^2=1$. This is equivalent to $\int_0^\pi u_{\ell \mathfrak{m},b}^\dagger u_{\ell \mathfrak{m},b}\,d\theta
=
\int_0^\pi v_{\ell \mathfrak{m},b}^\dagger v_{\ell \mathfrak{m},b}\,d\theta
=1$, where the superscript~\(\dagger\) denotes the Hermitian conjugation.
As expected, the above relations enforce $\ell \ge \max(2,|m|)$ for the paired sector. 

For $|m|\leq1$, there is also an unpaired dipole electromagnetic mode with
$\ell=1$, 
\begin{equation}
\label{eq:dipole-angular}
    \lambda_{\mathrm d}=-\frac{3M}{2Q},
\quad
F_{\mathrm d}=\sqrt{\sin\theta}\,{}_1S_{1\mathfrak{m}},
\quad
G_{\mathrm d}=0.
\end{equation}

Thus, for each $(\ell,m)$ the angular problem contains two paired
branches $b=\pm1$, each mixing spin-weight-$1$ and spin-weight-$2$
harmonics. These eigenvalues determine the corresponding radial master functions $R_{\ell m,b}^{(\varsigma)}$.
%

\subsection{Reconstruction of NP scalars in the static case}
\label{sec:NPreconstruction}

We reconstruct the rescaled NP scalars in~\ref{eq:rescaledpsi}. First, for each angular eigenvalue $\lambda_{\ell,b}$ determined in \ref{sec:AngularSolution}, we solve~\ref{eq:eqforRvarsigma} for the corresponding radial master function $R_{\ell m,b}^{(\varsigma)}$. The subsequent radial functions $(z_1^{(\varsigma)})_{\ell m,b}$ and $(z_2^{(\varsigma)})_{\ell m,b}$ then follow from~\ref{eq:eaforz12p} for $\varsigma=+1$ and~\ref{eq:eaforz12m} for $\varsigma=-1$. At fixed $m$, summing over $\ell$ and both branches $b=\pm1$ in~\ref{eq:zpzmDecomposition} yields $(z_+^{(\varsigma)},z_-^{(\varsigma)})^{\rm T}$. Finally, inverting the definitions of $z_{\pm}^{(\varsigma)}$ in~\ref{eq:zpzmp} and~\ref{eq:zpzmm} gives $(\psi_2,\psi_1,\hat{k},\hat{s})$ for $\varsigma=+1$ and $(\psi_{-2},\psi_{-1},\hat{n},\hat{\ell})$ for $\varsigma=-1$.

\subsubsection{Full expressions in the static case}
Henceforth, we focus on the static case~$(\omega=0)$. The pairs~$(\psi_2, \psi_1)$ and $(\psi_{-2}, \psi_{-1})$ are expressed in terms of the radial master functions~$R_{\ell m,b}^{(+1)}$ and $R_{\ell m,b}^{(-1)}$, respectively, and their derivatives. We shall use these reconstruction formulas in two different regimes. Near the future horizon, the frequency must be kept finite in order to distinguish the ingoing and outgoing solutions, and the static limit is taken only after imposing the physical boundary condition. At spatial infinity, instead, we can work directly in the static limit to identify the tidal source and response.

Specifically, using~\ref{eq:angulareigenfunctions} with~\ref{eq:solsFG}, we obtain the following relation between $\psi_{2,1}$ with $R_{\ell m,b}^{(+1)}$ and its first derivative, in the static limit, as,
\begin{widetext}
\begin{align}
&\psi_2=-\frac{i}{\bar{\rho} \Delta \left(M^2-Q^2\right)}
\nonumber
\\
&\times
\sum_{b =\pm1,  \ell m}\left[\left\{\frac{M \bar{\rho}-Q^2}{\Delta} R_{\ell m,b}^{(+1)}-\frac{2Q^2}{3M +2Q\lambda_{\ell ,b}} \frac{dR_{\ell m,b}^{(+1)}}{dr}\right\} C_{\ell ,b}^{\rm g} \,_2S_{\ell m}-i\frac{aQ}{\Delta} R_{\ell m,b}^{(+1)}  C_{\ell ,b}^{\rm e}  \sin\theta \,_1S_{\ell m}\right]e^{i m \widetilde{\phi}}\,,
\label{reconstructionpsi2}
\\
&\psi_1=\frac{iQ}{\bar{\rho} \Delta \left(M^2-Q^2\right)}
\nonumber
\\
&\times \sum_{b=\pm1, \ell m} \left[\left\{R_{\ell m,b}^{(+1)} -\frac{2\left(M \bar{\rho}-Q^2\right)}{3M+2Q\lambda_{\ell ,b}}\frac{dR_{\ell m,b}^{(+1)}}{dr} \right\} C_{\ell ,b}^{\rm e} \,_1S_{\ell m} -i \frac{2aQ}{3M+2Q\lambda_{\ell ,b}} \frac{dR_{\ell m,b}^{(+1)}}{dr} C_{\ell ,b}^{\rm g} \sin \theta \,_2S_{\ell m}\right]e^{i m \widetilde{\phi}}\,,
\label{reconstructionpsi1}
\end{align}
where we have ignored some constant phase factors. Here, $\widetilde{\phi}\equiv\phi+\bar r_*$, with $d\bar r_*/dr=a/\Delta$, is the ingoing azimuthal coordinate. The reconstructed NP scalars are thus expressed in ingoing coordinates $(v,\widetilde{\phi})$, where $v\equiv t+r_*$ and $dr_*/dr=(r^2+a^2)/\Delta$.\footnote{As discussed in~\ref{sec:DecoupledEquations}, all modes contain the factor $He^{-i\omega t+im\phi}$. The definition of $H$ in~\ref{eq:eaforz12m} gives
$He^{-i\omega t+im\phi}
=e^{-i\omega v+im\widetilde{\phi}}
e^{i\omega r_*(2r_+)-im\bar r_*(2r_+)}$,
where the last factor is a constant phase that can be absorbed into the mode amplitudes.}
Similarly, one can write the rescaled NP scalars for the negative-spin case in terms of $R^{(-1)}_{\ell m,b}$ and its first derivative as 
\begin{align}
&\psi_{-2}=\frac{iQ \Delta^2}{\bar{\rho}\left(M^2-Q^2\right)} \nonumber
\\
&\times \sum_{b=\pm 1,\ell m} \left[i\left\{ R_{\ell m,b}^{(-1)}+\frac{2\left(M \bar{\rho} -Q^2\right)}{3M -2Q \lambda_{\ell,b}}  \frac{dR_{\ell m,b}^{(-1)}}{dr}\right\}C_{\ell ,b}^{\rm g} \,_2S_{\ell \, -m} + \frac{2 a Q}{3M-2Q\lambda_{\ell,b}}\frac{dR_{\ell m,b}^{(-1)}}{dr} C_{\ell ,b}^{\rm e} \sin\theta \,_1S_{\ell \, -m}\right] e^{i m \tilde \phi}\,,\label{reconstructionpsim2}
\\
&\psi_{-1}=-\frac{i  \Delta^2}{\bar{\rho} \left(M^2-Q^2\right)} \nonumber
\\
&\times \sum_{b=\pm1,\ell m} \left[ - i\left\{\frac{M \bar{\rho}-Q^2}{\Delta} R_{\ell m,b}^{(-1)}+\frac{2Q^2}{3M -2Q \lambda_{\ell,b}}\frac{dR_{\ell m,b}^{(-1)}}{dr}\right\}  C_{\ell ,b}^{\rm e} \,_1S_{\ell \, -m}+ \frac{a Q}{\Delta} R_{ \ell m, b}^{(-1)} C_{\ell ,b}^{\rm g}\sin \theta\,_2S_{\ell \, -m}\right] e^{i m \tilde \phi}\,.\label{reconstructionpsim1}
\end{align}

\subsubsection{Near-horizon behavior and boundary conditions}
Here, we focus on the positive-spin system~$(\varsigma=+1)$ and derive the near-horizon behavior of the reconstructed scalars $\psi_2$ and $\psi_1$ using the master radial solution~$R^{(+1)}_{\ell m,b}$ given in~\ref{eq:general-solution} of the main text.
By expanding $\psi_{2}$ near the horizon, we obtain 
\begin{equation}
\label{limittohorizon2}
\begin{aligned}
\psi_2\Big|_{r\simeq r_+}
&\simeq
-\frac{i(M\bar{\rho}-Q^2)}
{\bar{\rho}\Delta^2(M^2-Q^2)}
\sum_{\ell m}
\left(
c_{1,+}C_{\ell,+}^{\rm g}
+c_{1,-}C_{\ell,-}^{\rm g}
\right)
\,{}_2S_{\ell m}e^{im\widetilde{\phi}}
\\[5pt]
&\quad
-\frac{aQ}{\bar{\rho}\Delta^2(M^2-Q^2)}
\sum_{\ell m}
\left(
c_{1,+}C_{\ell,+}^{\rm e}
+c_{1,-}C_{\ell,-}^{\rm e}
\right)
\sin\theta\,{}_1S_{\ell m}e^{im\widetilde{\phi}}
\\[9pt]
&\quad
-\frac{i}{\bar{\rho}(M^2-Q^2)(r_+-r_-)^4}
\sum_{\ell m}\Bigg[
\left(
M\bar{\rho}-Q^2
-\frac{2Q^2(1-\gamma)(r_+-r_-)}
{3M+2Q\lambda_{\ell,+}}
\right)c_{2,+}C_{\ell,+}^{\rm g}
\\
&\qquad\qquad
+\left(
M\bar{\rho}-Q^2
-\frac{2Q^2(1-\gamma)(r_+-r_-)}
{3M+2Q\lambda_{\ell,-}}
\right)c_{2,-}C_{\ell,-}^{\rm g}
\Bigg]
(-z)^{-p}\,{}_2S_{\ell m}e^{im\widetilde{\phi}}
\\[5pt]
&\quad
-\frac{aQ}{\bar{\rho}(M^2-Q^2)(r_+-r_-)^4}
\sum_{\ell m}
\left(
c_{2,+}C_{\ell,+}^{\rm e}
+c_{2,-}C_{\ell,-}^{\rm e}
\right)
(-z)^{-p}\sin\theta\,{}_1S_{\ell m}e^{im\widetilde{\phi}}\,.
\end{aligned}
\end{equation}
As evident, in the near-horizon regime, we have two branches, one behaving as $\Delta^{-2}$, while the other scaling as $z^{-p}$, where $p\equiv 2iam/(r_{+}-r_{-})$. 

Likewise, the near-horizon expansion of $\psi_1$ reads
\begin{equation}
\label{psi1nearhorizon}
\begin{aligned}
\psi_1\Big|_{r\simeq r_+}
&\simeq
\frac{iQ}{\bar{\rho}\Delta(M^2-Q^2)}
\sum_{\ell m}\Bigg[
c_{1,+}C_{\ell,+}^{\rm e}
+c_{1,-}C_{\ell,-}^{\rm e}
\\
&\qquad\qquad
+\frac{2(M\bar{\rho}-Q^2)\alpha_+\alpha_-}
{(r_+-r_-)\gamma}
\left(
\frac{c_{1,+}C_{\ell,+}^{\rm e}}
{3M+2Q\lambda_{\ell,+}}
+\frac{c_{1,-}C_{\ell,-}^{\rm e}}
{3M+2Q\lambda_{\ell,-}}
\right)
\Bigg]
\,{}_1S_{\ell m}e^{im\widetilde{\phi}}
\\[5pt]
&\quad
-\frac{2aQ^2}{\bar{\rho}\Delta(M^2-Q^2)}
\sum_{\ell m}
\frac{\alpha_+\alpha_-}{(r_+-r_-)\gamma}
\left(
\frac{c_{1,+}C_{\ell,+}^{\rm g}}
{3M+2Q\lambda_{\ell,+}}
+\frac{c_{1,-}C_{\ell,-}^{\rm g}}
{3M+2Q\lambda_{\ell,-}}
\right)
\sin\theta\,{}_2S_{\ell m}e^{im\widetilde{\phi}}
\\[9pt]
&\quad
-\frac{2iQ(M\bar{\rho}-Q^2)}
{\bar{\rho}(M^2-Q^2)(r_+-r_-)^3}
\sum_{\ell m}(1-\gamma)
\left(
\frac{c_{2,+}C_{\ell,+}^{\rm e}}
{3M+2Q\lambda_{\ell,+}}
+\frac{c_{2,-}C_{\ell,-}^{\rm e}}
{3M+2Q\lambda_{\ell,-}}
\right)
(-z)^{-p}\,{}_1S_{\ell m}e^{im\widetilde{\phi}}
\\[5pt]
&\quad
+\frac{2aQ^2}
{\bar{\rho}(M^2-Q^2)(r_+-r_-)^3}
\sum_{\ell m}(1-\gamma)
\left(
\frac{c_{2,+}C_{\ell,+}^{\rm g}}
{3M+2Q\lambda_{\ell,+}}
+\frac{c_{2,-}C_{\ell,-}^{\rm g}}
{3M+2Q\lambda_{\ell,-}}
\right)
(-z)^{-p}\sin\theta\,{}_2S_{\ell m}e^{im\widetilde{\phi}}\,.
\end{aligned}
\end{equation}

\end{widetext}
where, we again have two branches, one behaving as $\Delta^{-1}$, and the other scaling as $z^{-p}$. Thus, we obtain the following behaviour for the scalars $\psi_{1,2}$ in the near-horizon regime, 
\begin{align}\label{horizonobtained}
\psi_{s}\Big|_{r\simeq r_{+}}=\frac{\mathbb{A}_{s}}{\Delta^{s}}+\mathbb{B}_{s}z^{-p}\,;
\qquad 
s=1,2\,,
\end{align}
where the dependence on $e^{im\widetilde{\phi}}$ has been suppressed. As evident from \ref{limittohorizon2} and \ref{psi1nearhorizon}, the coefficients $\mathbb{A}_{1,2}$ depend only on linear combinations of $(c_{1+},c_{1-})$, while the coefficients $\mathbb{B}_{1,2}$ depend only on linear combinations of $(c_{2+},c_{2-})$. 

We identify the physical branch by comparing \ref{horizonobtained} with the zero-frequency limit of the near-horizon wave decomposition. By analogy with the Kerr analysis of Ref.~\cite{Teukolsky:1974yv}, we expect the finite-frequency KN scalars to have the leading behavior
\begin{align}
\psi_s\Big|_{r\simeq r_+}
\simeq \frac{{\cal A}_{\rm ingoing}^{(s)}}{\Delta^s}
+{\cal A}_{\rm outgoing}^{(s)}e^{2i\bar{\omega}r_*},
\qquad s=1,2,
\label{eq:horizon-wave-decomposition}
\end{align}
where the common factor $e^{-i\omega v+im\widetilde{\phi}}$ has been suppressed. The physical boundary condition at the future event horizon selects the purely ingoing solution, requiring
\begin{align}
{\cal A}_{\rm outgoing}^{(s)}=0.
\label{eq:ingoing-bc}
\end{align}
In the static limit, $\bar{\omega}\to-m\Omega_H$. Using the near-horizon behavior
$r_*=\frac{r_+^2+a^2}{r_+-r_-}\ln z+{\cal O}(1)$,
the outgoing factor becomes proportional to $z^{-p}$ at leading order, with a constant phase absorbed into its amplitude. The ingoing boundary condition therefore requires the coefficient of $z^{-p}$ in~\ref{horizonobtained} to vanish. Thus, $\mathbb{B}_{1,2}=0$, which implies
\begin{align}
\label{eq:regularity}
{c_{2,b}=0} \qquad b=\pm1\,,
\end{align}
as evident from \ref{limittohorizon2} and \ref{psi1nearhorizon}. 
This is the condition used in the main text. It selects the smooth radial branch, for which
$R_{\ell m,b}^{(+1)}=c_{1,b}+{\cal O}(z)$
near the future horizon. The same choice applies to the negative-spin system, so that both sectors are described by
\begin{align}
R_{\ell m,b}^{(\varsigma)}
=c_{1,b}\,{}_2F_1(\alpha_+,\alpha_-;\gamma;-z).
\label{eq:regular-static-radial}
\end{align}
Substituting this solution into~\ref{reconstructionpsi2}--\ref{reconstructionpsim1} gives the leading near-horizon behavior of each reconstructed mode:
\begin{align}
\psi_{2,\ell m b}\big|_{r\simeq r_+}
&\propto \Delta^{-2},
&\qquad
\psi_{1,\ell m b}\big|_{r\simeq r_+}
&\propto \Delta^{-1},
\label{eq:nearpsip}\\
\psi_{-2,\ell m b}\big|_{r\simeq r_+}
&\propto \Delta^2,
&\qquad
\psi_{-1,\ell m b}\big|_{r\simeq r_+}
&\propto \Delta.
\label{eq:nearpsim}
\end{align}
The positive-spin behavior in~\ref{eq:nearpsip} reproduces the static limit of the ingoing branch in~\ref{eq:horizon-wave-decomposition}. Both spin sectors are consistent with the horizon regularity condition stated in Ref.~\cite{Hintz:2026jiw}, which requires the appropriately rescaled fields to extend smoothly across the future horizon.\footnote{Ref.~\cite{Hintz:2026jiw} calls modes that are purely outgoing at future null infinity and purely ingoing at the future event horizon ``outgoing''. Its index $b\in\{2,1,-1,-2\}$ in Eq.~(2) labels the boost weight of the NP scalars and differs from our branch index $b=\pm1$.}
\subsubsection{Large-distance behavior: the Love matrix}

Now, we examine the large-distance asymptotic behavior of the reconstructed NP scalars and hence construct the associated Love matrix. In the static limit, at large distances~$r\gg r_+$, the rescaled NP quantities for  the positive-spin system behave as
\begin{widetext}
    \begin{equation}\label{eq:largepsipsolution}
    \begin{aligned}
      \psi_2 \big|_{r \gg r_+}\simeq & -\dfrac{i}{r^3\left(M^2-Q^2\right)} \sum_{b =\pm1, \ell m} \left[\frac{M}{r} R^{(+1)}_{\ell m,b} -\frac{2Q^2}{3M+2Q \lambda_{\ell,b}} \frac{d R^{(+1)}_{\ell m, b}}{dr}\right]  C_{\ell, b}^{\rm g}  \,_2S_{\ell m}  e^{im\widetilde{\phi}}\,,
      \\
    \psi_1 \big|_{r \gg r_+} \simeq & \dfrac{i Q }{r^3\left(M^2-Q^2\right)} \sum_{b =\pm1, \ell m}\left[ R^{(+1)}_{\ell m,b}-\frac{2M r}{3M+2Q \lambda_{\ell,b}} \frac{d R^{(+1)}_{\ell m,b}}{dr}\right]  C_{\ell, b}^{\rm e}  \,_1S_{\ell m} e^{im\widetilde{\phi}}\,.
\end{aligned}
\end{equation}
Likewise, the asymptotic behavior of the rescaled NP scalars for the negative-spin system is given by
\begin{equation}\label{eq:largepsimsolution}
\begin{aligned}
\psi_{-2}\big|_{r \gg r_+}\simeq & -\frac{Q r^3}{M^2-Q^2}\sum_{b=\pm 1
,\ell m} \left[R_{\ell m,b}^{(-1)} +\frac{2 M r}{3M-2Q\lambda_{\ell,b}} \frac{dR_{\ell m,b}^{(-1)}}{dr}  \right]C_{\ell,b}^{\rm g}  \,_2S_{\ell -m} e^{im\widetilde{\phi}}\,,
\\
\psi_{-1}\big|_{r\gg r_+}\simeq & -\frac{ r^2}{M^2-Q^2}\sum_{b=\pm 1,\ell m} \left[ M R_{\ell m,b}^{(-1)} +\frac{2 Q^2 r}{3M-2Q\lambda_{\ell,b}}\frac{d R_{\ell m,b}^{(-1)}}{dr}\right] C_{\ell ,b}^{\rm e} \,_1S_{\ell -m}  e^{im\widetilde{\phi}}\,.
\end{aligned}
\end{equation}
\end{widetext}
Here, the terms mixing different spin weights contribute only at subleading order in the large-distance expansion.

We now extract the components of the Love matrix from the asymptotic behavior of each mode with fixed~$(\ell, m)$ and for $\varsigma=+1$. The negative-spin case can be understood in a parallel manner. We begin by writing down the leading large-distance behavior of~\ref{eq:regular-static-radial}, which reads
\begin{equation}
\label{eq:Ratlarger2}
R^{(+1)}_{\ell m,b}
=\mathcal N_{\ell m,b} r^{\ell+2}
\Bigg[(1+\cdots)+\mathcal F_{\ell m}(1+\cdots)
\left(\frac{r_+}{r}\right)^{2\ell+1}
\Bigg]\,,
\end{equation}
where
\begin{equation}
\begin{aligned}
\mathcal F_{\ell m}
=
&-\frac{iam}{r_+}
\frac{(\ell+2)!(\ell-2)!}
{(2\ell+1)!(2\ell)!}\times\\
&\prod_{j=1}^{\ell}
\left[
j^2\left(1-\frac{r_-}{r_+}\right)^2
+\left(\frac{2am}{r_+}\right)^2
\right]\,,
\end{aligned}
\end{equation}
and 
\begin{equation}
\mathcal N_{\ell m,b}
=\frac{c_{1,b}}{(r_+-r_-)^{\ell+2}}
\frac{\Gamma(-1+p)\Gamma(2\ell+1)}
{\Gamma(\ell-1)\Gamma(\ell+1+p)}\,,
\end{equation}
where $p = 2i a m /(r_+-r_-)$. Substituting $R_{\ell m}^{(+1)}$ to \ref{eq:largepsipsolution}, we obtain
\begin{widetext}
\begin{equation}
\begin{aligned}
\psi_{2,\ell m}|_{r\gg r_+}\simeq&
-\frac{i}{M^2-Q^2} \sum_{b=\pm1}\mathcal N_{\ell m,b}
\\
&\times\left[
\left(M-\frac{2Q^2(\ell+2)}{3M+2Q\lambda_{\ell,b}}\right)r^{\ell-2}
+\mathcal F_{\ell m}r_+^{2\ell+1}
\left(M+\frac{2Q^2(\ell-1)}{3M+2Q\lambda_{\ell,b}}\right)r^{-\ell-3}
\right]
C_{\ell,b}^{\rm  g}\,{}_2S_{\ell m} e^{im\widetilde{\phi}},
\end{aligned}
\end{equation}
\begin{equation}
\begin{aligned}
\psi_{1,\ell m}|_{r\gg r_+}\simeq&
\frac{iQ}{M^2-Q^2}
\sum_{b=\pm1}\mathcal N_{\ell m,b}
\\
&\times\left[
\left(1-\frac{2M(\ell+2)}{3M+2Q\lambda_{\ell,b}}\right)r^{\ell-1}
+\mathcal F_{\ell m}r_+^{2\ell+1}
\left(1+\frac{2M(\ell-1)}{3M+2Q\lambda_{\ell,b}}\right)r^{-\ell-2}
\right]
C_{\ell,b}^{\rm e}\,{}_1S_{\ell m} e^{im\widetilde{\phi}}.
\end{aligned}
\end{equation}
These behaviors can be schematically expressed as
\begin{equation}
    \begin{aligned}
        \label{eq:reduced-radial-equation2}
    r^2 \psi_{2,\ell m} &\simeq \left[A^{\rm g}_{\ell m}r^{\ell}(1+\cdots) + B^{\rm g}_{\ell m}r^{-\ell -1}(1+\cdots)\right]{}_2S_{\ell m} e^{im\widetilde{\phi}}\,,
    \\
    r\psi_{1,\ell m}&\simeq \left[A^{\rm e}_{\ell m} r^\ell(1+\cdots) + B^{\rm e}_{\ell m} r^{-\ell-1}(1+\cdots)\right]{}_1S_{\ell m} e^{im\widetilde{\phi}},
    \end{aligned}
\end{equation}
where
\begin{align}
A_{\ell m}^{\rm  g}
&:
=-i\mathcal P_{\ell m}
\sum_{b=\pm1}\mathcal N_{\ell m,b}C_{\ell,b}^{\rm  g}
\left(
M-\frac{2Q^2(\ell+2)}{3M+2Q\lambda_{\ell,b}}
\right),
\\
B_{\ell m}^{\rm  g}
&:=-i\mathcal P_{\ell m}\mathcal F_{\ell m}r_+^{2\ell+1}
\sum_{b=\pm1}\mathcal N_{\ell m,b}C_{\ell,b}^{\rm  g}
\left(
M+\frac{2Q^2(\ell-1)}{3M+2Q\lambda_{\ell,b}}
\right),
\\
A_{\ell m}^{\rm  e}
&:=iQ\mathcal P_{\ell m}
\sum_{b=\pm1}\mathcal N_{\ell m,b}C_{\ell,b}^{\rm e}
\left(
1-\frac{2M(\ell+2)}{3M+2Q\lambda_{\ell,b}}
\right),
\\
B_{\ell m}^{\rm  e}
&:=iQ\mathcal P_{\ell m}\mathcal F_{\ell m}r_+^{2\ell+1}
\sum_{b=\pm1}\mathcal N_{\ell m,b}C_{\ell,b}^{\rm e}
\left(
1+\frac{2M(\ell-1)}{3M+2Q\lambda_{\ell,b}}
\right),\\
\mathcal P_{\ell m}&:=
\frac{1}{M^2-Q^2}.
\end{align}
\end{widetext}
Eliminating the arbitrary amplitudes $\mathcal N_{\ell m,b}$ in favor of the source coefficients $A^{g,e}_{\ell m}$, we obtain the susceptibility Love matrix presented in the main text.

Finally, we consider the dipole mode~($\ell=1$) introduced in~\ref{eq:dipole-angular}. Since spin-weight-$2$ spherical harmonics exist only for $\ell\ge2$, this mode has no dynamical gravitational sector. Its tidal response is purely electromagnetic and is characterized by the single coefficient
\begin{align}
    \frac{B^{\rm e}_{1m}}{A^{\rm e}_{1m}}
    =r_+^3\mathcal{F}_{1m}^{\rm ee}\,,
\end{align}
where $\mathcal{F}_{1m}^{\rm ee}$ was presented in the main text. The neutral limit~$Q\to 0$ reproduces the Kerr result~\cite{Chakraborty:2026qru}.

\subsubsection{Neutral limit~$Q\to 0$ for $\ell\geq2$}
We examine the reconstructed static NP scalars for $\ell\geq2$ in the neutral limit~$Q\to 0$. From ~\ref{reconstructionpsi2} to~\ref{reconstructionpsim1}, we obtain
\begin{equation}
\begin{aligned}
\label{eq:uncharged_positive_spin}
      \lim_{Q\to 0}\psi_2 = & -\dfrac{i}{\Delta^2 M} \sum_{\ell m} R^{(+1)}_{\ell m,+1} \,_2S_{\ell m} e^{i m \tilde{\phi}},\\
    \lim_{Q\to 0} (Q \psi_1)  =& \dfrac{i}{\Delta} \sum_{\ell m}   \frac{3}{\ell^2+\ell-2}\frac{d R^{(+1)}_{\ell m,-1}}{dr} \,_1S_{\ell m} e^{i m \tilde{\phi}},
\end{aligned}
\end{equation}
and
\begin{equation}
\begin{aligned}
\label{eq:uncharged_negative_spin}
\lim_{Q\to 0}\left(Q\psi_{-2}\right) = & \Delta^2 \sum_{\ell m} \frac{3}{\ell^2+\ell-2} \frac{dR_{\ell m,+1}^{(-1)}}{dr} \,_2S_{\ell -m} e^{i m \tilde{\phi}},\\
\lim_{Q\to 0}\psi_{-1} = & -\frac{\Delta}{M} \sum_{\ell m} R_{\ell m, -1}^{(-1)} \,_1S_{\ell -m} e^{i m \tilde{\phi}}.
\end{aligned}
\end{equation}
These expressions explicitly show that the $b=+1$ and $b=-1$ branches reduce to purely gravitational and purely electromagnetic perturbations, respectively. 
As discussed in~\ref{sec:nulltetradrotations}, the unscaled NP scalars admit finite uncharged limits provided $\psi_{\pm2}$ and $Q\psi_{\pm1}$ remain finite as $Q\to0$. For the positive-spin sector, \ref{eq:uncharged_positive_spin} shows that this condition is satisfied when the overall normalizations of $R_{\ell m,b}^{(+1)}$ are ${\cal O}(1)$ for both branches $b=\pm1$. For the negative-spin sector, \ref{eq:uncharged_negative_spin} instead requires the normalizations of $R_{\ell m,b}^{(-1)}$ to scale as ${\cal O}(Q)$ for $b=+1$ and ${\cal O}(Q^{-1})$ for $b=-1$.

\subsubsection{Static radial solutions: Case $am=0$ }
When $am=0$,~\ref{radial} must be solved separately. We first consider the axisymmetric case, $m=0$. Since the azimuthal number enters~\ref{radial} only through the combination $am$, this analysis also encompasses the general nonspinning case obtained by taking the $a\to0$ limit of the final result.

For $m=0$, the general solution can be written in terms of associated Legendre functions as
\begin{align}
    R_{\ell0,b}^{(\varsigma)}
    ={}&d_{1,b}\Delta^{\nu/2}
    P_{\ell}^{\nu}(1+2z)
    +d_{2,b}\Delta^{\nu/2}
    Q_{\ell}^{\nu}(1+2z),
    \label{solaxi}
\end{align}
where $P_{\ell}^{\nu}$ and $Q_{\ell}^{\nu}$ are associated Legendre functions of the first and second kind, respectively, and $2\nu=1+3\varsigma$.
Regularity at the outer horizon selects the first solution, implying $d_{2,b}=0$.

Using $\Delta=(r_+-r_-)^2z(1+z)$, the horizon-regular solution can equivalently be written in hypergeometric form as
\begin{align}
R_{\ell0,b}^{(\varsigma)}
={}&
\frac{
d_{1,b}\Gamma(\ell+\nu+1)(r_+-r_-)^{\nu}
}{
\Gamma(\ell-|\nu|+1)\Gamma(1+|\nu|)
}
[z(1+z)]^{(\nu+|\nu|)/2}
\nonumber\\
&\times
{}_2F_1\left(
|\nu|-\ell,\,
\ell+|\nu|+1;\,
1+|\nu|;\,
-z
\right).
\end{align}

We now focus on the positive-spin sector, $\varsigma=+1$, for which $\nu=2$. The regular solution reduces to
\begin{align}
R_{\ell0,b}^{(+1)}
={}&
\frac{d_{1,b}\Gamma(\ell+3)(r_+-r_-)^2}
{2\Gamma(\ell-1)}
z^2(1+z)^2
\nonumber\\
&\times
{}_2F_1(2-\ell,\ell+3;3;-z).
\end{align}
To extract the large-distance behavior, we analytically continue $\ell$ away from integer values and use the connection formula for the hypergeometric function,
\begin{widetext}
\begin{align}
{}_2F_1(2-\ell,\ell+3;3;-z)
={}&
\frac{2\Gamma(2\ell+1)}
{\Gamma(\ell+3)\Gamma(\ell+1)}
z^{\ell-2}
{}_2F_1\left(
2-\ell,-\ell;-2\ell;-\frac1z
\right)
\nonumber\\
&+
\frac{2\Gamma(-2\ell-1)}
{\Gamma(2-\ell)\Gamma(-\ell)}
z^{-\ell-3}
{}_2F_1\left(
\ell+3,\ell+1;2\ell+2;-\frac1z
\right).
\end{align}
\end{widetext}
Applying Euler's transformation to absorb the prefactor $z^2(1+z)^2$ and omitting an overall normalization yield
\begin{align}
R_{\ell0,b}^{(+1)}
\propto{}&
z^{\ell+2}\Bigg[
{}_2F_1\left(
-\ell-2,-\ell;-2\ell;-\frac1z
\right)
\nonumber\\
&+
\frac{
\Gamma(\ell+3)\Gamma(\ell+1)\Gamma(-2\ell-1)
}{
\Gamma(2\ell+1)\Gamma(2-\ell)\Gamma(-\ell)
}
z^{-2\ell-1}
\nonumber\\
&\hspace{1.5cm}\times
{}_2F_1\left(
\ell-1,\ell+1;2\ell+2;-\frac1z
\right)
\Bigg].
\end{align}
Therefore, at large distances, the leading-order terms for the source and response contributions take the form 
\begin{align}
R_{\ell0,b}^{(+1)}
=
\mathcal N_{\ell0,b}r^{\ell+2}
\left[
(1+\cdots)+\mathcal F_{\ell}(1+\cdots)
\left(\frac{r_+}{r}\right)^{2\ell+1}
\right],
\end{align}
with
\begin{align}
\mathcal F_{\ell}
=
\left(\frac{r_+-r_-}{r_+}\right)^{2\ell+1}
\frac{
\Gamma(-2\ell-1)\Gamma(\ell+3)\Gamma(\ell+1)
}{
\Gamma(2\ell+1)\Gamma(2-\ell)\Gamma(-\ell)
}.
\end{align}
Using the reflection formula for the Gamma function, the $\ell$-dependent singular factors combine as
\begin{align}
\mathcal F_{\ell}\propto
\frac{\sin^2(\pi\ell)}{\sin(2\pi\ell)}.
\end{align}
Taking the integer-$\ell$ limit after analytic continuation then yields
\begin{align}
\label{eq;Fellm0}
    \mathcal F_{\ell}=0,
    \qquad \ell =2,3,4 \ldots.
\end{align}
Note that this expression holds for $\ell\ge 2$. The dipole case~$(\ell=1)$ can be understood in the same manner as the~$am \neq 0$ case discussed in the main text. \ref{eq;Fellm0} shows that the axisymmetric static response vanishes. The same conclusion holds in the nonspinning limit $a\to0$ for arbitrary $m$, consistently with the $am\to0$ limit of the generic result.

\end{document}